\documentstyle[prl,aps,epsfig]{revtex}
\begin{document}
\draft \title{Vortex lattice structures of Sr$_2$RuO$_4$}
\author{D.F. Agterberg}
\address{Theoretische Physik, Eidgenossiche Technische Hochschule-
H\"{o}nggerberg, 8093 Z\"{u}rich, Switzerland}
\date{\today}
\maketitle
\begin{abstract}

The vortex lattice structures of Sr$_2$RuO$_4$ for the  
odd parity representations of the superconducting
state are examined for the magnetic field along the crystallographic
directions. Particular emphasis is placed upon the two dimensional
representation which is believed to be relevant to this material.
It is shown that when
the zero-field state breaks time reversal symmetry, there must exist
{\it two} superconducting  transitions when there is a finite field 
along a high symmetry direction in the basal plane. Also it is shown
that a square vortex lattice is expected 
when the field is along the $c$-axis. The orientation
of the square lattice with respect to the underlying ionic lattice yields
information as to which Ru 4$d$ orbitals are relevant to the
superconducting state.  
\end{abstract}

The oxide superconductor Sr$_2$RuO$_4$ is structurally 
similar to the high $T_c$ materials but differs 
markedly from 
the latter in its electronic structure \cite{mae94}.
In particular,  
the normal state near the superconducting transition 
of Sr$_2$RuO$_4$ is well described by
a quasi-2D Landau Fermi liquid \cite{mac296}. 
There now exists considerable evidence that the superconducting
state of Sr$_2$RuO$_4$ \cite{mae94} 
is not a conventional $s$-wave state. 
NQR measurements
show no indication of a Hebel-Slichter peak in $1/T_1T$ \cite{ish97},
$T_c$ is strongly suppressed by non-magnetic impurities \cite{mac98},
and tunneling experiments are inconsistent with $s$-wave
pairing \cite{jin98}. While these measurements demonstrate that
the superconducting state is non $s$-wave, they do not 
determine what pairing symmetry actually occurs in this material.
The determination of the pairing symmetry in unconventional 
superconductors is a notoriously difficult problem and 
theoretical insight provides a useful guide.
The observations that
the Fermi liquid 
corrections due to  
electron correlations are similar in magnitude to those found in
superfluid $^3$He and that closely related ruthenates are 
itinerant ferromagnets   
have led to the proposal that the 
superconducting state in Sr$_2$RuO$_4$ is of odd-parity \cite{ric95}.
Even with this insight there still remain five odd-parity states
that have different symmetry -  all of which have a nodeless gap and
therefore similar thermodynamic
properties \cite{ric95}. Recently, $\mu$SR measurements 
indicate that a spontaneous internal magnetization begins to develop
at $T_c$ \cite{luk98}. The most natural interpretation 
of this magnetization is that
the superconducting state {\it breaks} time reversal symmetry (${\cal T}$).
This places a strong constraint on the pairing symmetry
in Sr$_2$RuO$_4$ since it implies that the superconducting 
order parameter must have more than one component \cite{sig91}.
Of the possible representations (REPS) of the $D_{4h}$ point group 
only the two dimensional (2D) $\Gamma_{5u}$ and $\Gamma_{5g}$ 
REPS exhibit this property. Of these two the
$\Gamma_{5u}$ REP is the most likely to occur in Sr$_2$RuO$_4$
due to arguments of Ref.~\cite{ric95} and 
the quasi-2D nature of the electronic properties.  
The order parameter in this case
has two components $({\eta_1,\eta_2})$
that share the same rotation-inversion  symmetry
properties as $(k_x,k_y)$ \cite{sig91}. The broken ${\cal T}$ state would
then correspond to $(\eta_1,\eta_2)\propto(1,i)$.

I investigate within Ginzburg Landau (GL) theory the vortex 
lattice structures expected for 
the odd-parity REPS of the superconducting state; focusing mainly 
 on the $\Gamma_{5u}$ REP. 
It is initialy shown that a general consequence 
of the broken ${\cal T}$ state described above   
is that in a finite magnetic field
oriented along a high symmetry direction in the basal plane there
will exist a {\it second} superconducting transition in the mixed state 
as temperature is
reduced. The observation of such a transition would provide
very strong evidence that the order parameter belongs to the
$\Gamma_{5u}$ REP. 
The form of the vortex lattice for a field along
the $c$-axis is then investigated  for both the one dimensional
(1D) and the 2D $\Gamma_{5u}$ REPS
of the
superconducting state.  
It is shown that a square vortex lattice is expected to appear for all the 
REPS,  however  
observable differences exist 
between the 1D and the 2D REPS.
Finally, within the recently proposed 
model of orbital dependent superconductivity of Sr$_2$RuO$_4$ \cite{agt97} 
it is also shown that the orientation of square vortex lattice 
with respect to the underlying crystal lattice dictates    
which of the Ru $4d$ orbitals give rise
to the superconducting state.

To demonstrate the presence of the two superconducting  
transitions described above consider 
the magnetic field along the $\hat{x}$ direction 
($x$ is chosen to be along the basal plane main crystal axis)
 and a homogeneous zero-field state $(\eta_1,\eta_2)\propto(1,i)$.
 In general the presence
of a magnetic field along the $\hat{x}$ direction breaks the
degeneracy of the $(\eta_1,\eta_2)$ components, so that only
one component will order  
at the upper critical field [{\it e.g.} $(\eta_1,\eta_2)\propto(0,1)$]. 
As has been
shown for type II superconductors with a single component
the vortex lattice is hexagonal at $T_c$ and the order parameter
solution is independent of $x$ so that $\sigma_x$
(a reflection about the $\hat{x}$ direction) 
is a symmetry operation of the $(\eta_1,
\eta_2)\propto (0,1)$ vortex phase \cite{luk95}.
Now consider the zero-field phase $(\eta_1,\eta_2)\propto(1,i)$, 
$\sigma_x$ transforms $(1,i)$ to $(-1,i)\ne e^{i\phi}(1,i)$ 
where $\phi$ is phase factor. This implies that $\sigma_x$ is {\it not}
a symmetry operator of the zero-field phase. It follows that there
must exist a transition in the finite field phase at 
which $\eta_1$ becomes non-zero. Similar arguments hold 
for the field along any
of the other three crystallographic directions in the basal 
plane. Evidence for this transition may 
already exist in the ac magnetic susceptibility measurements
of Yoshida {\it et. al.} \cite{yos96}. They observed a second peak in the
imaginary part of the magnetic susceptibility only when the
flux lines were parallel to the basal plane. 

For a more detailed analysis 
consider the following dimensionless GL free energy density for the 
$\Gamma_{5u}$ REP 
\begin{eqnarray} 
f=&-|\vec{\eta}|^2+|\vec{\eta}|^4/2+\beta_2(\eta_1\eta_2^*-\eta_2\eta_1^*)^2/2
+\beta_3|\eta_1|^2|\eta_2|^2 +|D_x\eta_1|^2+|D_y\eta_2|^2\label{eq1}\\
&+\kappa_2(|D_y\eta_1|^2+ 
|D_x\eta_2|^2)+ \kappa_5(|D_z\eta_1|^2+|D_z\eta_2|^2)
\nonumber \\ & +\kappa_3[(D_x\eta_1)(D_y\eta_2)^*+h.c.]+
\kappa_4[(D_y\eta_1)(D_x\eta_2)^*+h.c.]
+h^2.
\nonumber
\end{eqnarray}
where $h=\nabla\times {\bf A}$, 
$D_{\nu}=\nabla_{\nu}/\kappa-iA_{\nu}$, 
$f$ is in units $B_c^2/(4\pi)$, lengths are 
in units $\lambda=[\hbar^2 c^2 \beta_1/(16 e^2 \kappa_1 \alpha \pi)]^{1/2}$,
$h$ is in units $\sqrt{2}B_c=\Phi_0/(4\pi\lambda\xi)^{1/2}$,
$\alpha=\alpha_0(T-T_c)$, $\xi=(\kappa_1/\alpha)^{1/2}$, 
and $\kappa=\lambda/\xi$. Note that $\lambda,\xi$, $B_c$ and
$\kappa$ are simply convenient choices and do not 
correspond to measured values of these parameters. 
A thorough analysis of Eq.~\ref{eq1} is difficult due
to the unknown phenomenological parameters $\beta_2,\beta_3,
\kappa_2,\kappa_3$, and $\kappa_4$. To simplify the analysis these
parameters are determined in the weak-coupling, clean-limit for
an arbitrary Fermi surface. 
Taking for the $\Gamma_{5u}$ REP the gap function  
described by the pseudo-spin-pairing gap matrix:
$\hat{\Delta}=i[\eta_1 v_x/\sqrt{\langle v_x^2\rangle}+
\eta_2v_y/\sqrt{\langle v_x^2\rangle}]\sigma_z \sigma_y$
where the brackets $\langle \rangle$ denote an average over the Fermi surface
and $\sigma_i$ are the Pauli matrices, it is  found that 
$\beta_2=\kappa_2=\kappa_3=\kappa_4=\gamma$ and   
$\beta_3=3\gamma-1$ where $\gamma=\langle v_x^2v_y^2 \rangle /
\langle v_x^4 \rangle$. 
Note that $0\le\gamma\le1$ and that $\gamma=1/3$ for a cylindrical 
or spherical Fermi surface. These parameters agree with 
the cylindrical Fermi surface results of Ref.~\cite{zhu97}.
It is easy to verify that in zero-field  
$(\eta_1,\eta_2)\propto(1,i)$  is the stable ground
state for all $\gamma$.

It is informative to determine the values of $\gamma$
that are relevant to Sr$_2$RuO$_4$. LDA band structure calculations
\cite{ogu95,sin95} reveal that the
density of states
near the Fermi surface are due mainly to the four Ru $4d$ electrons
in the $t_{2g}$ orbitals.
There is a strong hybridization of these orbitals with the O
$2p$ orbitals giving rise to antibonding $\pi^*$ bands. The resulting
bands have three quasi-2D Fermi surface sheets labeled $\alpha,\beta,$
and $\tilde{\gamma}$ (see Ref. \cite{mac296}). The $\alpha$ and $\beta$ sheets
consist of $\{xz,yz\}$ Wannier functions and the
$\tilde{\gamma}$ sheet of $xy$ Wannier functions. 
In general $\gamma$
is not given by a simple average over all the sheets
of the Fermi surface.  A knowledge of the pair scattering
amplitude on each sheet and between the sheets is required
to determine $\gamma$ \cite{agt97,maz97}.
Recently, to account for
the large residual density of states observed in the superconducting
state, it has been proposed that either the 
$xy$ or the $\{xz,yz\}$ Wannier functions exhibit
superconducting order \cite{agt97}. This model implies that 
that there are two possible values of $\gamma$; one for the $\tilde{\gamma}$
sheet ($\gamma_{xy}$) and one for an average over the $\{\alpha,\beta\}$
sheets ($\gamma_{xz,yz}$). A tight binding model  
indicates $\gamma_{xy}=0.67$ and $\gamma_{xz,yz}=0.11$ 
\cite{agt98}. 
These values are sensitive to changes in the parameters of the tight
binding model, however the qualitative result that $\gamma_{xy}>1/3$ and
$\gamma_{xz,yz}<1/3$ is robust. Physically $\gamma_{xy}>1/3$
because of the proximity of the $\tilde{\gamma}$ Fermi surface sheet
to a Van Hove singularity and $\gamma_{xz,yz}<1/3$ due to quasi
1D nature of the $\{\alpha,\beta\}$ surfaces \cite{ogu95,sin95}.

\begin{figure}
\epsfxsize=160mm
\epsffile{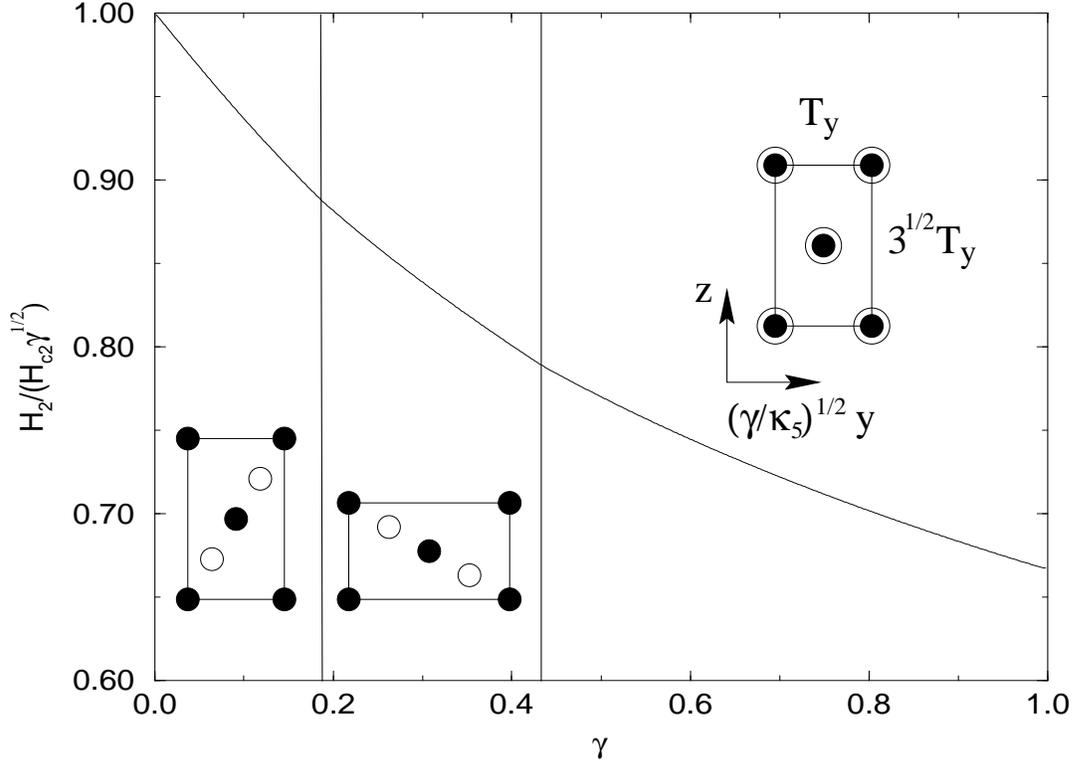}
\vglue 0.1 cm
\caption[*]{
The ratio of the two transition lines for the field along
$\hat{x}$ as a function of $\gamma$. The open (closed) circles
correspond to the zeroes of the $\eta_2$ ($\eta_1$) lattice.
The vertical lines separate
regions where the depicted vortex lattice structures are favored.
For all three lattice structures the $y$ and $z$ axes have the same
orientation and the dimensions of the rectangular cell are the same.}
\label{fig1}
\end{figure}

Following Burlachov \cite{bur85} for the 
solution of upper critical field $H_{c_2}^{ab}$ for 
the field in the basal plane, 
the vector potential
is taken to be ${\bf A}=Hz(\sin\theta,-\cos\theta,0)$  ($\theta$ is
the angle the applied magnetic field makes with
the $\hat{x}$ direction). After setting the
component of ${\bf D}$ along the field to be zero it is found that 
$H_{c_2}^{ab}(\theta)=\kappa/(\kappa_5\lambda(\theta)/2)^{1/2}$
where
$\lambda(\theta)=1+\gamma-[(1-\gamma)^2-(1+\gamma)(1-3\gamma)
\sin^2 2\theta ]^{1/2}$.
A measurement of the temperature independent four-fold 
anisotropy in $H_{c_2}^{ab}$ thus determines $\gamma$. 
To determine the field at which the second
transition discussed above occurs consider the magnetic field 
along the $\hat{x}$ direction. 
The free energy of Eq.~\ref{eq1} is then similar to that studied in
UPt$_3$ \cite{gar94,joy91,luk95} and since Sr$_2$RuO$_4$ is a strong
type II superconductor with a GL parameter of 31 for the field in the basal
plane \cite{yos96}
the procedure of Garg and Chen
\cite{gar94} to study the second transition can be applied here.
At $H_{c_2}^{ab}$ $\eta_1$ orders and
and the vortex lattice solution is given by \cite{luk95,gar94,joy91} 
\begin{equation}
\eta_1=\sum_nc_ne^{i nqz}e^{-(\kappa_5/\gamma)^{1/2}\kappa 
H[y-qn/(\kappa H)]^2/2} \label{eqeta1}.
\end{equation}  
where $c_n=e^{in^2\pi/2}$ and $q$ has the two possible
values $q_1^2=\sqrt{3}H\kappa \pi (\gamma/\kappa_5)^{1/2}$
or $q_2^2=H\kappa \pi (\gamma/\kappa_5)^{1/2}/\sqrt{3}$
(these two solutions are degenerate in energy).
At the second transition the
$\eta_2$ component becomes non-zero. As
is discussed in Refs. \cite{gar94,joy91} the solution 
for $\eta_2$ corresponds to a lattice  
that is displaced relative to that of $\eta_1$ by ${\bf d}=(\bar{y},\bar{z})$. Accordingly, the field at which the second transition occurs 
is found
by substituting 
\begin{equation}
\eta_2=ir\sum_nc_ne^{i(nq+\kappa H \bar{y})(z-\bar{z})}e^{-
\sqrt{\kappa_5}\kappa 
H[y-\bar{y}-qn/(\kappa H)^2]/2}
\end{equation}  
and Eq.~\ref{eqeta1} into the free energy,
minimizing with respect to the displacement vector ${\bf d}$, 
and determining when the coefficient of $r^2$ 
becomes zero. This yields for the
ratio of the second transition ($H_2$) to the upper critical field 
\begin{equation}
\frac{H_2}{H_{c_2}^{ab}}=\gamma^{1/2}\frac{\beta_A-\gamma(2S_1-|S_2|)
_{min}}
{\beta_A-\gamma^{3/2}(2S_1-|S_2|)_{min}}\label{eq4}
\end{equation}
where $\beta_A=1.1596$, $S_1=\overline{|\eta_1|^2|\eta_2|^2}/
(\overline{|\eta_1|^2}\hphantom{a}\overline{|\eta_2|^2})$,
$S_2=\overline{(\eta_1\eta_2^*)^2}/(\overline{|\eta_1|^2}
\hphantom{a}\overline{|\eta_2|^2})$, the over-bar denotes a spatial average,
and the subscript $min$
means take the minimum value with 
respect to ${\bf d}$ and with respect to $q=q_1$ or $q=q_2$.
The numerical solution of Eq.~\ref{eq4} is shown in Fig.~\ref{fig1}.
Three vortex lattice configurations are found to be stable
as a function of $\gamma$ (depicted in Fig.~\ref{fig1}).
For $0<\gamma<0.187$ $q=q_2$ and ${\bf d}=(T_y,T_z)/4$
($T_y$ and $T_z$ are the translation vectors of the centered
rectangular cell for the $\eta_1$ lattice), 
for $0.187<\gamma<0.433$ $q=q_1$ and ${\bf d}=(T_y,T_z)/4$, and
for $0.433<\gamma<1$ $q=q_2$ and ${\bf d}=0$.
For the field along $\hat{x}\pm\hat{y}$ the ratio $H_2/H_{c_2}$ is given by
Eq.~\ref{eq4} with $\gamma$ replaced by $(1-\gamma)/(1+3\gamma)$.
As a consequence the form of the vortex lattice will depend on the
field direction.
As has been discussed in detail in Ref.~\cite{joy97} 
the shape of the vortex lattice unit cell for $H<H_2$ will be
strongly field dependent.

\begin{figure}
\epsfxsize=120mm
\epsffile{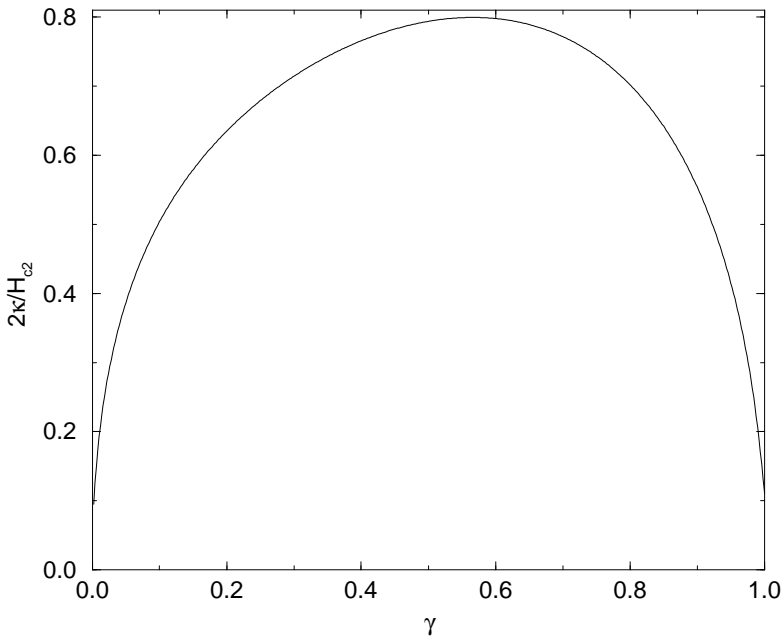}
\vglue 0.1 cm
\caption[*]{
The inverse of the upper critical field as a function of
$\gamma$ for the field applied along the $c$-axis.}
\label{fig2}
\end{figure}

Consider the magnetic field oriented along the $c$-axis. 
Setting $D_z=0$ writing $\Pi_+=\kappa(iD_x+D_y)/(2H)$,
$\Pi_-=\kappa(iD_x-D_y)/(2H)$, $\eta_+=(\eta_x+i\eta_y)/\sqrt{2}$,
and $\eta_-=(\eta_x-i\eta_y)/\sqrt{2}$,  minimizing the quadratic
portion of Eq.~\ref{eq1} with respect to $\eta_+$ and $\eta_-$ 
yields
\begin{equation}
2\kappa \pmatrix{\eta_+\cr \eta_-}=H\pmatrix{(1+\gamma)(1+2N)
&(1+\gamma)\Pi_+^2+(1-3\gamma)\Pi_-^2\cr (1+\gamma)\Pi_-^2+(1-3\gamma)\Pi_+^2&
(1+\gamma)(1+2N)}\pmatrix{\eta_+\cr \eta_-}\label{eq5}.
\end{equation}
where $N=\Pi_+\Pi_-$. The maximum value of $H$ that allows a non-zero
solution for $(\eta_+,\eta_-)$ yields the upper critical field $H_{c_2}^c$.
For $\gamma\ne 1/3$ $H_{c_2}^c$ must be found numerically (note that for
$\gamma=1/3$ the solution can be found analytically \cite{zhi89,sig91}). 
Expanding 
$(\eta_+,\eta_-)$ in terms of the eigenstates of $N$ (Landau levels) 
up to $N=32$ 
and diagonalizing the
resulting matrix yields the result for $H_{c_2}^c(\gamma)$ shown
in Fig.~\ref{fig2}.  The solution for the form of the vortex lattice
represents a complex problem due to presence of many Landau levels
in the solution of $(\eta_+,\eta_-)$ and the 
weak type II nature of Sr$_2$RuO$_4$ for the field along
the $c$ axis (Ref.~\cite{yos96} indicates $\kappa\approx 1.2$).
Here I present results that are strictly valid in the large $\kappa$ 
limit and leave the treatment for general $\kappa$  
to a later publication (a perturbative calculation indicates
 that the qualitative results are unchanged for
$\kappa=1.2$) \cite{agt98}.    
The form of the eigenstate
of the $H_{c_2}^c$ solution is found to be 
$\eta_+({\bf r})=\sum_{n\ge 0} a_{4n+2} \phi_{4n+2}({\bf r})$ and  
$\eta_-({\bf r})=\sum_{n\ge 0} a_{4n} \phi_{4n}({\bf r})$ where 
$\phi_n({\bf r})=
\sum_m c_me^{iqm\tilde{y}}2^{-n/2}H_n(\tilde{x}-qm/(\kappa H))e^{-\kappa 
H(\tilde{x}-qm/(\kappa H))^2/2}/(n!)^{1/2}$
where the coefficients $a_n$ are real, 
$(\tilde{x},\tilde{y})$ is the vector $(x,y)$ rotated by an angle 
$\tilde{\theta}$ about the $z$ axis and 
$H_n(x)$ represent Hermite polynomials. In the large $\kappa$ 
limit the
form of the vortex lattice is found by minimizing $\beta=\overline{f_4}
/(\overline{|\eta|^2})^2$ ($f_4$ is the quartic part
of Eq.~\ref{eq1}) with respect to the 
coefficients $c_n$, $q$, and $\tilde{\theta}$. It is assumed that 
$c_n=c_{n+2}$. This restricts the vortex lattice structures to
be centered rectangular with a short axis $L_y=2\pi/q$ and a long 
axis $L_x=2q/(\kappa H)$. The ratio $t=L_x/L_y$ is $\sqrt{3}$ for a
hexagonal vortex lattice and is 1 for a square vortex lattice.
I further restrict the analysis to
the two orientations $\tilde{\theta}=\{0,\pi/4\}$ since these correspond
to aligning one of the vortex lattice axes with one of the 
high symmetry directions in the basal plane.
Remarkably, the treatment
of the many Landau levels in the solution of $\eta_+$ and
$\eta_-$ becomes numerically straightforward when $\beta$ is 
expressed as a sum over the reciprocal lattice given by 
${\bf l}=\hat{x}l_12\pi/L_x+\hat{y}l_22\pi/L_y$ 
\cite{agt98} (see also Ref.~\cite{fra96}).  
It is found that $\beta$ is minimized for $c_n=e^{in^2\pi/2}$
and that the values of $t$ and $\tilde{\theta}$ depend upon
$\gamma$. For $\gamma\le 1/3$ ($\gamma\ge 1/3$) $\tilde{\theta}=0$ 
($\tilde{\theta}=\pi/4$) and $t$ 
varies continuously from $\sqrt{3}$ to 1 as $\gamma$ decreases (increases)
from $1/3$ to $1/3-0.0050$ ($1/3+0.0050$). For $\gamma<1/3-0.0050$ and 
$\gamma>1/3+0.0050$ the minimum $\beta$ corresponds to $t=1$. 
This implies that for $\gamma_{xz,yz}$ 
a square vortex lattice rotated $\pi/4$ about
the $c$ axis from the crystal lattice is expected and
for $\gamma_{xy}$ a square vortex lattice 
that is aligned with the underlying crystal lattice is expected
near $H_{c_2}^c$.
Note the appearance of the square vortex lattice correlates
with an anisotropy in $H_{c_2}^{ab}$ 
of $|1-H_{c_2}^{ab}(\theta=0)/H_{c_2}^{ab}(\theta=\pi/4)|>0.01$.

The recent observation of a square vortex lattice in Sr$_2$RuO$_4$
\cite{for98} makes it
of interest to compare the above behavior to that expected
for the 1D REPS of $D_{4h}$. It is well
known that for single component order parameters
non-local corrections to the standard GL theory stabilize a 
square vortex lattice \cite{tak71,aff96,dew97}. In particular the 
the following non-local term will stabilize the
square lattice
\begin{equation}
\epsilon [|(D_x^2-D_y^2)\psi|^2-|(D_xD_y+D_yD_x)\psi|^2].
\end{equation}
Treating this term as a perturbation to the GL free energy leads to 
$\psi=\phi_0-\tilde{\epsilon}\phi_4$ where 
$\tilde{\epsilon}=\sqrt{6}\epsilon H/\kappa$ ($\kappa$ 
is the GL parameter)
near $H_{c_2}^c$. As $\tilde{\epsilon}$ increases (note $\tilde{\epsilon}=0$ at $T_c$)
the vortex  lattice continuously distorts from hexagonal to square
\cite{tak71,aff96,dew97}
until the square vortex lattice is stable for $|\tilde{\epsilon}|>0.024$. 
The sign of $\epsilon$ determines the
orientation of the vortex lattice; for $\epsilon>0$ the vortex lattice
is aligned with the underlying lattice while for $\epsilon<0$ the 
lattice is rotated $\pi/4$ with respect to the underlying crystal
lattice \cite{tak71,dew97}. The sign of $\epsilon$ has been determined within 
a weak coupling clean limit approximation 
for the 1D odd-parity REPS. For the  
$A_{1u}$ REP using $\hat{\Delta}=\psi(\hat{x}v_x/\sqrt{\langle v_x^2 \rangle}+\hat{y}v_y/\sqrt{\langle v_x^2 \rangle})\cdot
\vec{\sigma} i\sigma_y$ 
the sign of $\epsilon$ is determined by the sign of 
$3\langle(v_x^2+v_y^2)v_x^2v_y^2\rangle/\langle(v_x^2+v_y^2)v_x^4\rangle-1$.
Using a form for $\hat{\Delta}$ that is analogous to that used for the
$A_{1u}$ REP, the same result is found for all the 1D odd-parity REPS.
This implies that the final orientation of the square vortex
lattice for the $1D$ REPS is the same as that found for the $2D$
REP for superconducting order in the $xy$ or the $\{xz,yz\}$ orbitals.
The behavior of the vortex lattice for
the 1D REPS as a function of $\tilde{\epsilon}$ is very
similar to that for the 2D REP as
a function of $\gamma$. An observable difference
between the 2D and the 1D REPS is that for the
2D REP the vortex lattice remains square up to $T_c$ 
while for the 1D REPS the vortex lattice is
hexagonal at $T_c$. Also, the  GL theories for the 1D and the 2D REPS 
predict a four-fold anisotropy in $H_{c_2}^{ab}$
but this anisotropy vanishes at $T_c$ for the 1D REPS and
does not vanish at $T_c$ for the 2D REP.    

In conclusion I have examined GL models for the
odd-parity REPS of the superconducting state for Sr$_2$RuO$_4$.
It was found that 
if the zero-field ground state breaks ${\cal T}$ symmetry 
(the 2D REP) then there should exist  a second 
transition in the mixed state when the magnetic field
is applied along a high symmetry direction in the basal
plane. It was also shown that when the field is along the $c$-axis there
will be a square vortex lattice for all the possible odd parity 
superconducting states.

I acknowledge support from the Natural Sciences
and Engineering Research Council of Canada and  the Zentrum for 
Theoretische Physik. I wish to thank E.M. Forgan, G.M. Luke, A. Mackenzie, 
Y. Maeno, T.M. Rice, and M. Sigrist for useful discussions.


\begin{references}
\bibitem{mae94} Y. Maeno, H. Hashimoto, K.Yoshida,S.Nishizaki,
T. Fujita, J.G. Bednorz, and F. Lichtenberg, Nature {\bf 372}, 532 (1994).

\bibitem{mac296} A.P Mackenzie, S.R. Julian, A.J. Diver, G.G. Lonzarich,
Y. Maeno, S.Nishizaki, and T. Fujita, Phys. Rev. Lett. {\bf 76}, 3786 (1996).


\bibitem{ish97} K. Ishida, Y.Kitaoka, K. Asayama, S.Ikeda, and T. Fujita,
Phys. Rev. B {\bf 56}, 505 (1997).

\bibitem{mac98} A.P. Mackenzie, R.K.W. Haselwimmer, A.W. Tyler, G.G Lonzarich,
Y. Mori, S. Nishizaki, and Y. Maeno, Phys. Rev. Lett. {\bf 80}, 161 (1998).

\bibitem{jin98} R. Jin, Yu. Zadorozhny, Y. Liu, Y. Mori, Y. Maeno,
  D.G. Schlom, and F. Lichtenberg, J. Chem. Phys. of Solids, in press.



\bibitem{ric95} T.M. Rice and M. Sigrist, J. Phys.: Condens. Matter {\bf 7},
L643 (1995).


\bibitem{luk98} G.M. Luke, Y. Fudamoto, K.M. Kojima, M.I. Larkin, B. Nachumi,
Y.J. Uemura, Y. Maeno, Z. Mao, Y. Mori, and H. Nakamura, to be published. 

\bibitem{sig91} M. Sigrist and K. Ueda, Rev. Mod. Phys. {\bf 63}, 239 (1991).



\bibitem{agt97} D.F. Agterberg, T.M. Rice, and M. Sigrist, Phys. Rev.
Lett. {\bf 78},3374 (1997).

\bibitem{luk95} I.A. Luk'yanchuk and M.E. Zhitomirsky, Supercond. 
Rev. {\bf 1}, 207 (1995).

\bibitem{yos96} K. Yoshida, Y. Maeno, S. Nishizaki, and T. Fujita, 
J. Phys. Soc. Jpn. {\bf 65}, 2220 (1996).


\bibitem{zhu97} J.X. Zhu, C.S. Ting, J.L. Shen, and Z.D. Wang, Phys. Rev. B
{\bf 56}, 14 093 (1997).

\bibitem{ogu95} T. Oguchi, Phys. Rev. B {\bf 51}, 1385 (1995).

\bibitem{sin95} D.J. Singh, Phys. Rev. B {\bf 52}, 1358 (1995).

\bibitem{maz97} I.I. Mazin and D.J. Singh, Phys. Rev. Lett. {\bf 79},
733 (1997).

\bibitem{agt98} D.F. Agterberg, in preparation.

\bibitem{bur85} L.I. Burlachkov, Sov. Phys. JEPT {\bf 62}, 800 (1985).

\bibitem{gar94} A. Garg and D.C. Chen, Phys. Rev. B {\bf 49}, 479 (1994).

\bibitem{joy91} R. Joynt, Europhys. Lett. {\bf 16}, 289 (1991).

\bibitem{joy97} R. Joynt, Phys. Rev. Lett. {\bf 78}, 3185 (1997).

\bibitem{zhi89} M.E. Zhitomirskii, JEPT Lett. {\bf 49}, 378 (1989).


\bibitem{fra96} M. Franz, C. Kallin, P.I. Soininen, A.J. Berlinsky,
and A.L. Fetter, Phys. Rev. B {\bf 53}, 5795 (1996).

\bibitem{for98} E.M. Forgan, {\it et. al.}, to be published.


\bibitem{tak71} K. Takanaka, Prog. Theor. Phys. {\bf 46}, 1301 (1971).

\bibitem{aff96} I. Affleck, M. Franz, and M.H.S. Amin,
Phys. Rev. B {\bf 55}, R704 (1996).

\bibitem{dew97} Y. De Wilde, M. Iavarone, U. Welp, V. Metlushko,
  A.E. Koshelev, I. Aranson, G.W. Crabtree, and P.C. Canfield, Phys. 
Rev. Lett. {\bf 78}, 4273 (1997).

\end{references}
\end{document}